\documentclass[11pt,twocolumn,aip,apl,reprint]{revtex4-1}

\usepackage{braket}
    
\usepackage{graphicx,amsmath,amssymb,braket}

\begin{document}

\title{Exact solution for driven oscillations in plasmonic field-effect transistors}
\author{D. Svintsov}
\affiliation{Laboratory of 2d Materials' Optoelectronics, Moscow Institute of Physics and Technology, Dolgoprudny 141700, Russia}
\affiliation{Institute of Physics and Technology, Russian Academy of Science, Moscow 117218, Russia}

\begin{abstract}
High-mobility field effect transistors can serve as resonant detectors of terahertz radiation due to excitation of plasmons in the channel. The modeling of these devices previously relied either on approximate techniques, or complex full-wave simulations. In this paper, we obtain an {\it exact} solution for driven electrical oscillations in plasmonic field-effect transistor with realistic contact geometry. The obtained solution highlights the importance of evanescent plasma waves excited near the contacts, which qualitatively modify the detector responsivity spectra. We derive the boundary condition on the ac floating electrodes of plasmonic FET which interpolates between open-circuit (Dyakonov-Shur) and short-circuit (clamped voltage) boundary conditions. In both limits, the FET photovoltage possesses resonant fringes, however, the absolute value of voltage is greater in the open-circuit regime.
\end{abstract}

\maketitle

\section{Introduction}
Antenna-coupled field-effect transistors (FET) have proven to be among most efficient detectors of THz radiation benefiting from room-temperature operation, high-speed response and compact size~\cite{Knap_PhysicsandImaging,Knap_Resonant,Peralta2002,Muravev2012,Shur_homodyne,Regensburger_detection}. The operating principle of such detectors is based on excitation of plasma waves in the FET channel~\cite{Dyakonov_detection_mixing} with subsequent rectification by various nonlinearities~\cite{RyzhiiShottky,cai2014sensitive,Giliberti-hydro}. Independently of the rectification mechanism, the response of FET-based detectors is strongly enhanced when incoming radiation induces plasma resonance in the channel. The proof-of-concept for resonant plasmonic THz detection was demonstrated with FETs based on two-dimensional electrons in III-V heterostructures~\cite{Knap_Resonant,Peralta2002,Muravev2012,Giliberti-hydro} and, more recently, with graphene~\cite{koppens2017acoustic}. Apart from enhancement of responsivity under resonant conditions, tuning of plasmon resonance by the gate voltage opens up the prospects for compact on-chip spectrometers~\cite{Muravev2012}.

The design of plasmonic on-chip THz spectrometers requires the ability to predict the frequencies of plasma resonance and electric field distributions in the channel. While the properties of plasma waves in extended two-dimensional systems are well-understood~\cite{Stern_Polarization,Chaplik1972}, the behavior of plasmons in confined systems is much more complex. Qualitatively, confinement of two-dimensional electron system (2DES) by source and drain contacts leads to the quantization of plasmon wave vector $q$, which becomes inversely proportional to the channel length $L$, $q_n\sim \pi n/L$~\cite{dyakonov1993shallow}. However, exact formulation of quantization conditions is hardly possible. The reason lies in strong non-locality effects; in other words, the wave equation for two-dimensional plasmons represents an integral equation, not differential one. Consequently, not only plane waves are excited in confined structures, but also the evancesscent waves near the contacts~\cite{Sydoruk_TransmissionGUG,Sydoruk_PlasmonsCoupled}. From the first glance, the complexity of plasmonic response in confined 2DES makes full electromagnetic simulations~\cite{khorrami2014,Popov_tuning} an only tool to predict their resonant properties.

In this paper, we obtain an {\it exact} solution for antenna-driven plasma oscillations in 2DES-based transistor with realistic contact geometry. The distribution of electric field in the channel is expressed as a rapidly converging series of quantized bulk plasmon modes plus the evanescent contribution with simple analytical structure. The solution allows us to recognize the general features of gate coupling to various plasmon modes, and to analyze previously used simplified approaches to plasmonic FET modelling. We find that commonly accepted quasi-optical approximation~\cite{dyakonov_ungated,satou_plasma_transit,partly_gated} breaks down very early as the gate-to-channel separation approaches plasmon wavelength. This occurs due to the above-mentioned neglect of evanescent waves.

We highlight the new boundary condition at antenna-decoupled electrodes, which interpolates between zero-current (Dyakonov-Shur) and fixed-potential conditions.  We use the obtained field distributions to simulate the responsivity spectra of plasmonic FETs assuming the hydrodynamic nonlinearities as main rectification mechanism. The real responsivity spectrum can be qualitatively different from commonly abovementioned analytical models. The rectified signal demonstrates resonant features for both open- and short-circuit conditions, but the open-circuit condition leads to higher responsivity.

\section{Plasmonic FET model and boundary conditions}

The plasmonic FET geometry is shown in Fig. 1: the 2DES is located at distance $d$ below the gate and confined by source and drain metal contacts, which we model as perfectly conducting walls. The THz antenna feeds the voltage $V_{g}e^{-i\omega t}$ between source and gate terminals, while the ac potential of the drain is floating. In real experiments~\cite{Knap_PhysicsandImaging,Knap_Resonant,bandurin2018,Regensburger_detection}, the dc signal (photocurrent or photovoltage $\bar V$) is measured between source and drain. The calculation of rectified signal requires the linear-response ac electric potential in the channel $\varphi_{\omega}(x,z)$ as an elementary building block. The latter can be found from the solution of the Poisson's equation
\begin{equation}
\label{Poisson}
\frac{\partial^2\varphi_\omega}{\partial x^2} + \frac{\partial^2\varphi_\omega}{\partial z^2} = \frac{4\pi e}{\varepsilon} n_\omega (x) \delta (z + d),
\end{equation}
where the right-hand side is responsible for induced electrons with density $n_\omega$ in the 2DES, $\varepsilon$ is the background dielectric constant.
 
\begin{figure}[t]
	\includegraphics[width=0.9\linewidth]{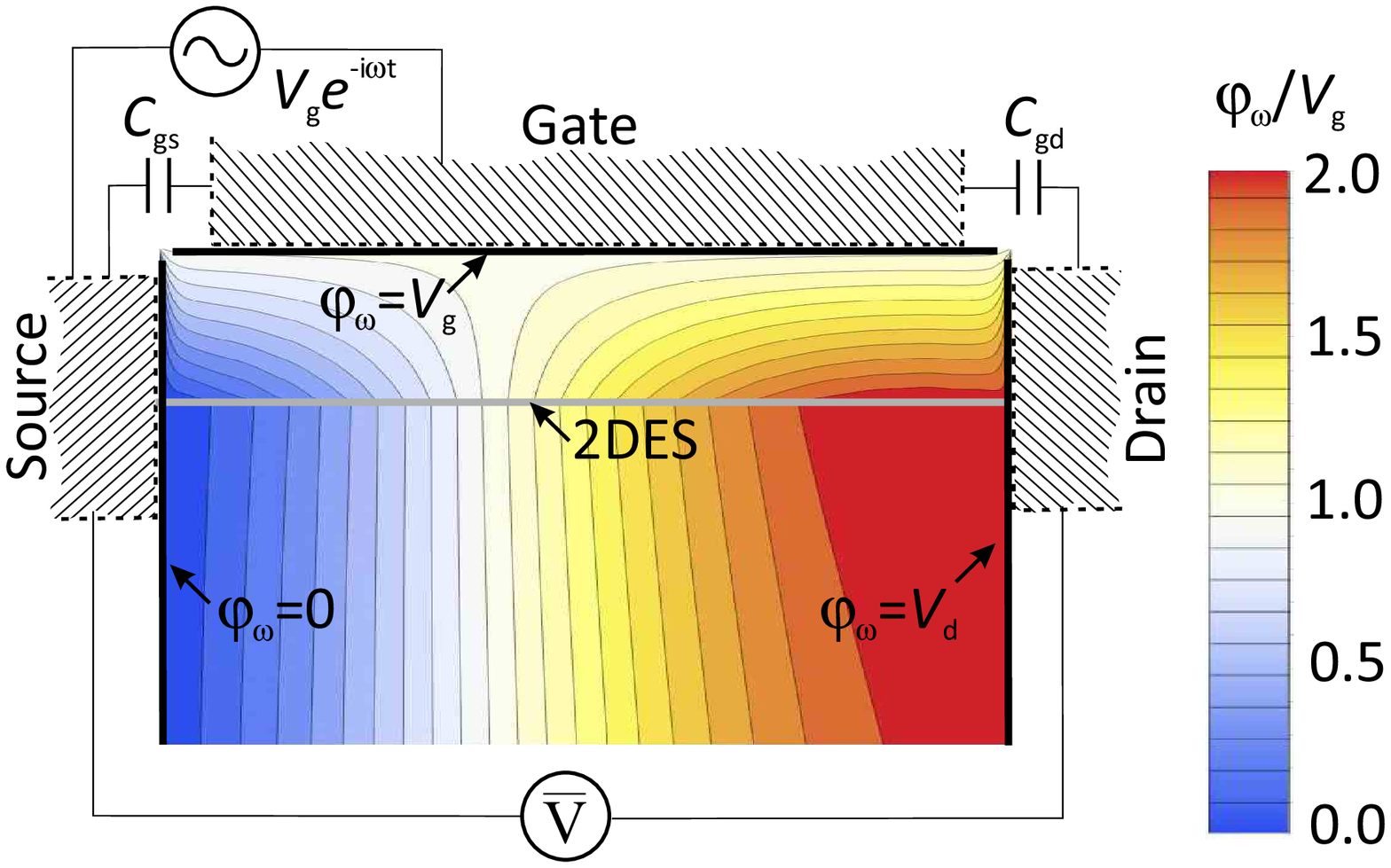}
	\caption{\label{Scheme} 
	Schematic of a field-effect transistor with two-dimensional channel overlaid with color map of electric potential distribution under plasmonic resonance.The signal from THz antenna $V_g e^{-i\omega t}$ is fed between source and gate, rectified dc voltage $\bar V$ is measured between source and drain. The ac potential $V_d$ is floating and determined from boundary condition (\ref{First-BC}). Hatched boxes denote the contacts of a realistic transistor. In calculation, they are mimicked by semi-infinite equipotential lines of source [$x=0$, $z \in(-\infty,0) $], drain [$x=L$, $z \in(-\infty,0) $], and finite-length gate [$x\in (0,L)$, $z = 0$]. $C_{gs}$ and $C_{gd}$ are parasitic gate-to-source and gate-to-drain capacitances
	}
\end{figure}

The boundary conditions at source and gate dictated by the THz antenna can be chosen as:
\begin{gather}
\varphi(0,z) = 0,\\
\varphi(x,0) = V_{g}.
\end{gather} 

The boundary condition at the floating electrode where the dc signal is read out requires special care. A key observation to set this condition is large inductive impedance of massive metallic contacts at terahertz frequencies. Indeed, already the inductance of an $l=1$ cm long and $w=0.1$ mm wide metal wire is ${\mathcal L}\approx 4l \ln(4l/w)\approx 10$ nH, which translates into reactive impedance $Z_L = - i \omega L\sim -80i$ k$\Omega$ at 1 THz frequency. This much exceeds the typical resistivity of two-dimensional systems. Therefore, the FET electrodes are ac-decoupled from the ''outer world'', and the drain potential is determined only by the transistor interior and its immediate vicinity, but not by the loading details.

Large reactive impedance of FET wiring implies no ac current in external circuit. In other words, all current flowing into the drain is charging the drain-to-gate ($C_{gd}$) and drain-to-source ($C_{sd}$) capacitance:
\begin{equation}
\label{First-BC}
{i_d} \left[\varphi \right]  = \frac{d}{dt}\left[C_{gd}(V_g - V_d)+C_{sd}(V_s - V_d)\right].
\end{equation}
Due to the variations of current density along the channel $j(x)$, the current $i_d$ is not merely the local current in the FET channel $j(L)$. Instead, this current is given by Shockley-Ramo theorem~\cite{Shockley,Ramo}:
\begin{equation}
{i_d} \left[\varphi \right] = \frac{1}{L}\int_0^L{dx j(x) g(x)},
\end{equation}
where $g(x)$ is the electric field in the channel if the drain electrode is at unit potential and all other electrodes grounded.

Large distance between source and drain electrodes allows us to neglect the capacitance $C_{sd}$ compared to $C_{gd}$ in (\ref{First-BC}). Assuming also the local relation between current density and electric field in 2DES $j(x)=-\sigma_{2d} d\varphi/dx$, we arrive at the boundary condition at the drain electrode
\begin{equation}
\label{BC}
V_d = V_g - \frac{\sigma_{2d}}{i\omega C_{gd}}\int_0^L{dx  g(x) \frac{d\varphi}{dx} }.
\end{equation}
For small separations between gate and drain (large capacitance), these electrodes are virtually short-circuited. For large separations, the channel current averaged with Shockley-Ramo theory should approach zero. Zero average current does not at all imply zero ac field at the drain. We shall see that this correspondence holds only for very small gate-to-channel separations when the drain potential in the channel is strongly screened.

\section{Exact solution of coupled Poisson and transport equations} 
To describe antenna-driven plasmons, we supplement the Poisson's equation (\ref{Poisson}) with current continuity equation and Ohm's law. This allows us to express induced density as
\begin{equation}
\label{Density}
n_\omega = \frac{en_0}{m \omega (\omega + i \tau^{-1})}\frac{\partial^2\varphi_\omega}{\partial x^2},
\end{equation}
where we have assumed the dynamics of 2d electrons to obey the Drude conductivity $\sigma_{2d} = n_0 e^2\tau_p/m/(1-i\omega\tau_p)$, where $n_0$ is the steady-state carrier density, $\tau_p$ is the momentum relaxation time, and $m$ is the electron effective mass (we set $m=0.067m_0$ in the following calculations). In the course of derivation it will become clear that the problem can be solved with any model for frequency-dependent conductivity.

To proceed further, we set extra boundary conditions on electron density at the contacts as $n_{\omega}|_{x=0} = n_{\omega}|_{x=L} =0$, which are justified by high contact doping and strong scattering therein. We note that solutions of coupled system (\ref{Poisson})-(\ref{Density}) with {\it zero} boundary conditions at all terminals are standing waves in the $x$-direction and linear combinations of exponentials in the vertical direction
\begin{equation}
\varphi_n = \phi_n(x) \psi_n(z) \equiv \sin q_n x \left\{ \begin{aligned}
& -\frac{\sinh {{q}_{n}}z}{\sinh {{q}_{n}}d},\,\,z>-d \\ 
& {{e}^{{{q}_{n}}\left( z+d \right)}},\,\,z\le -d \\ 
\end{aligned} \right.
\end{equation}
where $q_n = \pi n/L$. The full electric potential has also the contributions from gate and drain contacts, $\phi_d(x,z)$ and $\phi_g(x,z)$, that are localized at short distances and can be considered as evanescent solutions. These contact contributions are conveniently expressed via dimensionless form-factors $\phi_d(x,y) =V_d f_d (x,y)$, $\phi_g (x,y) = V_{g} f_g(x,y)$, where
\begin{gather}
\label{DrainPot}
f_d=\frac{1}{2}+\frac{1}{\pi }\arctan \left[ \frac{1+\cos \frac{\pi x}{L} \cos \frac{\pi y}{L} }{\sin \frac{\pi x}{L} \sin \frac{\pi y}{L} } \right],\\
\label{GatePot}
f_g=1+\frac{2}{\pi }\arctan \left[ \frac{\sinh \frac{\pi y}{L} }{\sin \frac{\pi x}{L} } \right].
\end{gather}
The full solution is sought for in the form
\begin{equation}
\phi(x,z) = \sum\limits_{n=0}^{\infty}{c_n \phi_n(x,y)}+V_d f_d(x,y)+V_g f_g(x,y).
\end{equation}
We introduce the full solution to the Poisson's equation, integrate it across the 2d layer, and use the orthogonality of $\phi_n$ to obtain the expressions for coefficients
\begin{equation}
{c_n}=\gamma_n(\omega) \left[ {V_g}{{\tilde f}_d}\left( q_n \right)+{V_g}{f_g}\left( {q_n} \right) \right],
\end{equation}
where $\gamma_n(\omega)$ can be viewed as dynamic response function
\begin{equation}
\gamma_n = \left[\frac{\omega(\omega+i\tau_p^{-1})}{\omega_p^2(q_n)} - 1\right]^{-1},
\end{equation}
$\omega_{p}(q)$ is the 2d plasma frequency  
\begin{equation}
\label{Plasma_freq}
\omega _{p}^{2}\left( q \right)=\frac{2\pi n{{e}^{2}}q/\varepsilon m}{1+\coth \left( qd/2 \right)},
\end{equation} 
the coefficients ${\tilde{f}}_d \left( q_n \right)$ and ${f}_g \left( q_n \right) $ are the coupling constants between plasmon potential and drain/gate potential. In fact, they are sine-Fourier transforms of (\ref{DrainPot}) and (\ref{GatePot}):
\begin{gather}
{\tilde{f}}_d \left( q_n \right) = (-1)^n \frac{2e^{- q_n d}}{q_n L} ,\\
{{f}_{g}}\left( {{q}_{n}} \right)=\left\{ \begin{aligned}
& \frac{4{{e}^{-{{q}_{n}}d}}}{{{q}_{n}}L},\,\,n\,\,{\rm odd}, \\ 
& 0\,\,{\rm otherwise}. \\ 
\end{aligned} \right..
\end{gather}

The closure of the solution is achieved once we use the boundary condition (\ref{BC}) to find the floating drain potential. This results in
\begin{equation}
\frac{V_d}{V_g}=\frac{1+\frac{\alpha_\omega}{2\pi }\left[ 2\coth \frac{\pi d}{L}+\sum\limits_{n}{\tilde{f}_{d}^{2}\left( {{q}_{n}} \right){{\left( q_nL \right)}^2}{\gamma_n}} \right]}{1+\frac{\alpha_\omega}{2\pi }\left[ \frac{4}{\sinh \frac{2 \pi d}{L}}-\sum\limits_{n}{{{\tilde{f}}_d}\left( q_n \right){{f}_{g}}\left( q_n \right){{\left( q_nL \right)}^2}{\gamma_n}} \right]}
\end{equation}
where
\begin{equation}
\alpha_\omega = \frac{\pi \sigma_{2d}}{i\omega C_{gd}L} \equiv \frac{\omega_{\rm eff}^{2}}{\omega(\omega + i \tau_p^{-1})}
\end{equation} 
is the ratio of capacitive impedance at the drain end of FET and the channel impedance. We readily observe that at high frequencies, the coefficient $\alpha_\omega$ tends to zero, which implies that gate and drain are effectively short-circuited. The microscopic evaluation of frequency $\omega^2_{\rm eff} = \pi n_0 e^2/C_{gd} m L$ that determines the crossover between open- and short-circuit conditions requires the knowledge of gate-to-drain capacitance. The latter appears to be infinite in the considered simplified geometry and will be left hitherto as a free parameter.

\begin{figure}[t]
	\includegraphics[width=0.9\linewidth]{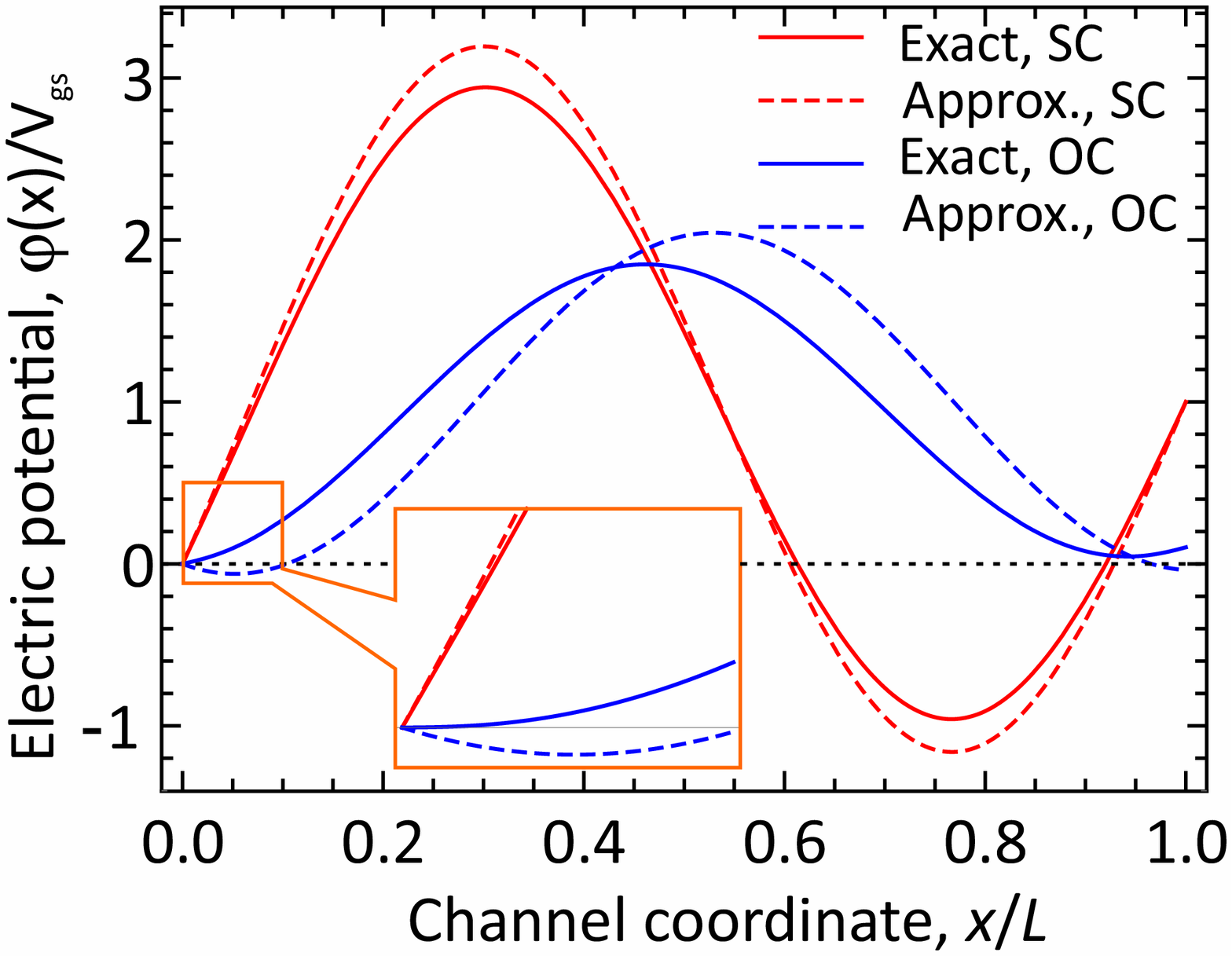}
	\caption{\label{Comparison_phi} 
		Comparison of electric potential distributions in a plasmonic FET obtained with exact solution (solid lines) and quasi-optical approximation (dashed lines). Short-circuit boundary condition at the drain ($|\alpha_\omega|\gg 1$, $V_d\rightarrow V_{gs}$) is shown in red, open-circuit boundary condition ($|\alpha_\omega|\ll 1$, $i_d \rightarrow 0$) is shown in blue. Channel length $L=1$ $\mu$m, carrier density $n_0=10^{12}$ cm$^{-2}$, momentum relaxation time $\tau_p = 2.5$ ps, frequency $\omega/2\pi = 1$ THz.
	}
\end{figure}

\section{Analysis of the solution}
The exact solution of forced oscillation problem in plasmonic FETs can be now analyzed and compared to the established analytical approaches to THz FET modeling. The most common one is the quasi-optical approximation, where the electric potential $\phi_{qo}$ is sought for as a linear combination of forward and backward waves, while the wave amplitudes are obtained from boundary conditions~\cite{dyakonov_ungated,satou_plasma_transit,partly_gated}. This procedure leads us to approximate solution
\begin{equation}
\phi_{qo} =V_{g}\left[1 - \frac{e^{i q_{p}(\omega) x} + r_\omega e^{- i q_{p}(\omega) (x - 2L)}}{1+r_\omega e^{2iq_{p}(\omega) L}}\right],
\end{equation}
where 
\begin{equation}
r_\omega = \frac{i \alpha_\omega q_{p}L /\pi - 1}{i \alpha_\omega q_{p}L /\pi + 1}
\end{equation}
is the wave reflection coefficient from the ac floating FET terminal, and $q_p(\omega)$ is the wave vector of 2d plasmon at frequency $\omega$, given by the inverse of Eq. (\ref{Plasma_freq}).

There are two major assumptions in the derivation of quasi-optical approximation. First, only one spatial harmonic with wave vector $q_{p}(\omega)$ contributes to channel potential, while all others (including evanescent modes) are ignored. Second, the reflection coefficient from the drain terminal relies on a local relation between current in drain circuit and channel current, $i_d = -\sigma_{2d} \left.d\varphi/dx\right|_{x=L}$. Both assumptions should work fine if gate-to-channel separation is below the plasmon wavelength. This is illustrated in Fig.~\ref{Comparison_phi}, where we compare the exact solutions (solid lines) with approximate ones (dashed lines) for short-circuit ($\alpha = 0$) and open-circuit ($\alpha \gg 1$) boundary conditions at the drain.

Even the ''average'' agreement of potential distributions in exact and simplified models does not yet imply the agreement of calculated FET responsivities as THz detectors. As the effects of THz rectification mostly occur at metal-semiconductor contacts~\cite{Muravev2012,RyzhiiShottky}, the responsivity is sensitive to the details of contact electric field. These electric fields are quite different in quasi-optical model and exact solution, as shown in the inset.

To be precise, we focus on THz rectification mechanism exploiting the intrinsic hydrodynamic nonlinearities of electron fluid~\footnote{The rectification by Schottky junctions at the contacts differs only with a prefactor of square brackets. In the case of junction rectification, it equals $|\sigma_{2d}|^2 (d^2 i/dV^2)/(di/dV)^2$ instead of $(e/m)/(\omega^2 + \tau_p^{-2})$}. The voltage responsivity $R_V$ of the FET-detector in this case is proportional to the difference of squared ac electric fields at contacts~\cite{Dyakonov_detection_mixing}
\begin{equation}
R_V =\frac{e/m}{\omega^2 + \tau^{-2}_p}\frac{Z_a}{V_g^2}\left[ |E_{x\omega}|^2_{x=0} - |E_{x\omega}|^2_{x=L}\right],
\end{equation}
here $Z_a$ is the antenna radiative resistance.

\begin{figure}[t]
	\includegraphics[width=0.75\linewidth]{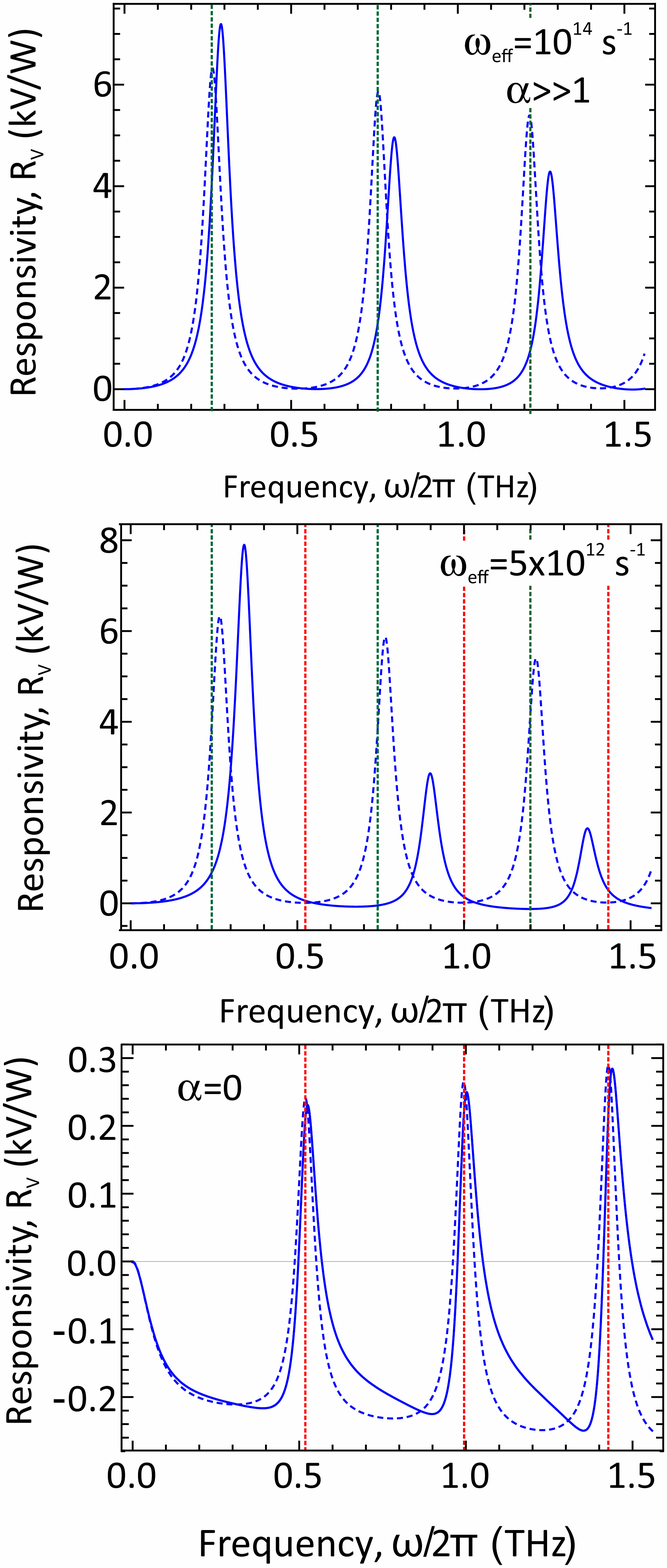}
	\caption{\label{RV} 
	Calculated responsivity of plasmonic THz detector assuming hydrodynamic nonlinearity as the main rectification mechanism for three values of effective frequency $\omega_{\rm eff}$ governing the boundary condition at the drain. (A) $\omega_{\rm eff} = 10^{14}$ s$^{-1}$, $\alpha \gg 1$ - zero current condition (B) $\omega_{\rm eff} = 5\times 10^{12}$ s$^{-1}$ -- intermediate boundary condition (C) $\omega_{\rm eff} = 0$ -- drain is short-circuited to gate. All parameters (except frequency $\omega$) are the same as in fig.~\ref{Comparison_phi}. Solid line shows the exact solution, dashed line - quasi-optical approximation. Green vertical lines shows the expected position of resonance for quarter-wavelength quantization condition $q_p L = \pi (k+1/2)$, red vertical lines -- for half-wavelength condition $q_p L = \pi k$.
	}
\end{figure}
The calculated ''hydrodynamic responsivity'' is shown in Fig. 3 for gate-to-channel separation $d=30$ nm and various boundary conditions. For open-circuit conditions, $\alpha \gg 1$, the responsivity can be reliably calculated with quasi-optical approximation. But as the capacitor impedance $Z_C=i/\omega C_{gd}$ goes down, the two solutions become different, bot in magnitude and positions of resonant peaks. In this regime, the potential of floating drain strongly depends on incoming current, which should be accurately calculated with Shockley-Ramo theorem. When $Z_C$ goes to zero, the potential of drain electrode is effectively pinned to the gate potential, and the positions of resonances agree in the two models. However, away from resonances the picture is quite different. The reason is strong overlap of evanescent drain field with higher-order modes in the channel which contribute to the potential distribution.
\begin{figure}[t]
	\includegraphics[width=0.8\linewidth]{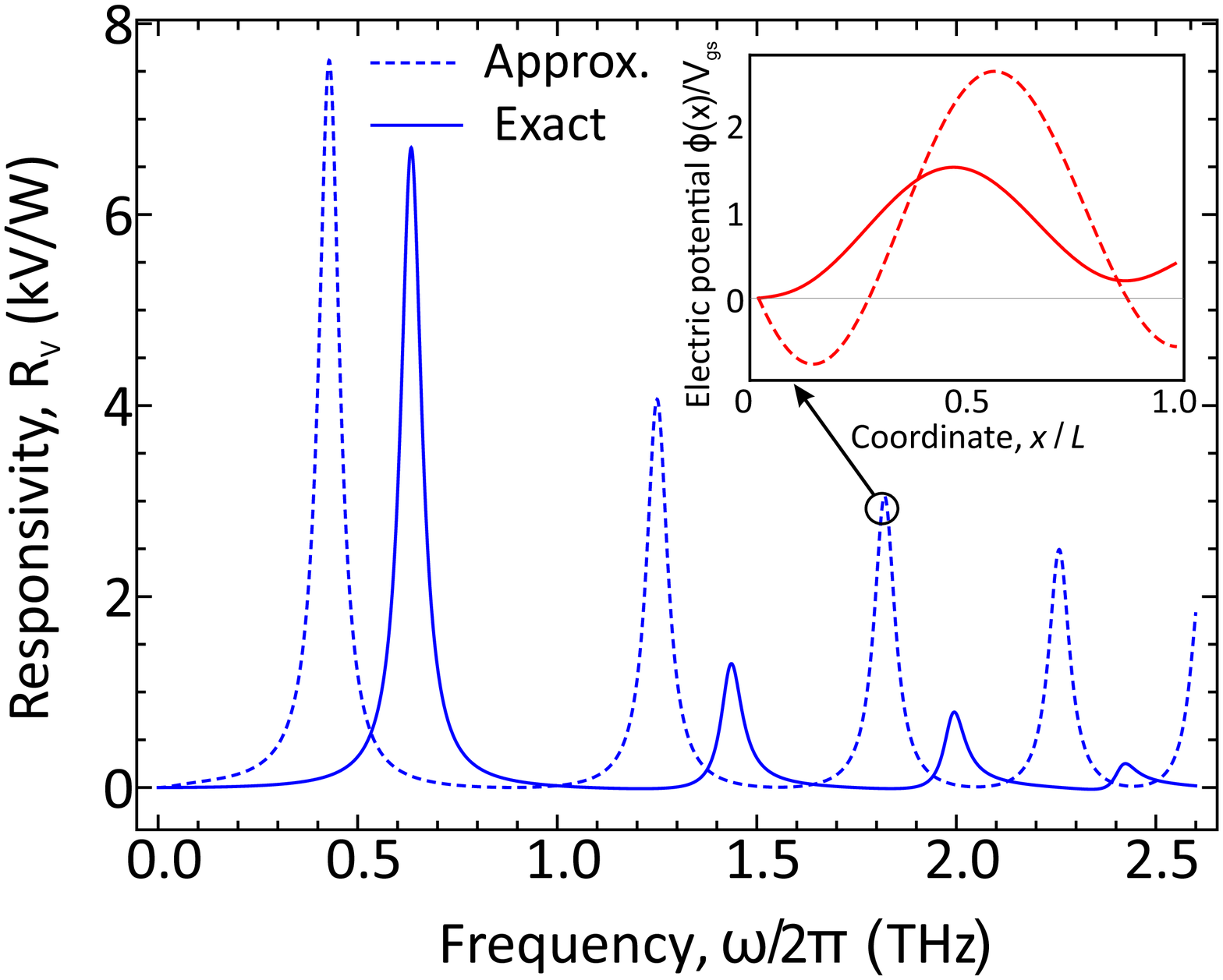}
	\caption{\label{RV_large_d} 
		Calculated THz FET responsivity assuming hydrodynamic nonlinearity as the main rectification mechanism for relatively large gate-to-channel separation $d=120$ nm and open-circuit boundary condition $|\alpha_\omega| \gg 1$. Solid line shows the exact solution, dashed line shows the result of quasi-optical approximation. Inset shows the potential distributions at frequency $\omega/2\pi = 1.8$ THz, which appear to be very different in the two approaches
	}
\end{figure}
As the gate-to-channel separation is further increased, the quasi-optical approximation breaks down very quickly, as illustrated in Fig. 4 for $d=120$ nm. Despite the channel is formally long, $L/d\gg 1$, the two models yield qualitatively different behaviour of resonant responsivity vs harmonic number. Namely, the exact solution shows that peak responsivity drops exponentially with $n$, $R_n \propto e^{-2 q_n d}$. The underlying reason is the evanescent character of the gate field which is formally manifested in exponential decay of its far Fourier harmonics.

\section{Conclusions}
We have presented an exact solution for driven electrical oscillations in plasmonic field-effect transistor with two-dimensional high-mobility channel. The considered device geometry and biasing conditions closely follow those used in contemporary resonant detectors: the high-frequency voltage is fed from THz antenna between source and gate terminals, while the rectified voltage is read out between source and drain~\cite{Knap_PhysicsandImaging,Knap_Resonant,bandurin2018,Regensburger_detection}. As expected, the amplitude of electric field in the channel is increased under the conditions of plasma resonance. However, these calculated profiles of electric potential and responsivity differ substantially in the exact solution and established analytical models using the plane-wave expansion of electric potential. One crucial aspect not taken into account by such models is the emergence of evanescent fields near the contacts. This is explicitly taken into account in exact solution. Another aspect is the floating electric potential of electrodes disconnected from antenna which should be accurately determined using the Shockley-Ramo theorem. We have found that at low frequencies and/or small gate-to-drain capacitance, the plasma resonances occur when the channel length accommodates odd number of plasmon quarter-wavelengths (known as Dyakonov-Shur condition). At high frequencies and/or large gate-to-drain capacitance, the resonances occur when the channel hosts an integer of half-wavelengths. The excitation of high-order harmonics by the oscillating gate potential is exponentially suppressed as the plasma wavelength approaches gate-to-channel separation. 

The model can be readily extended to non-Drude models of 2DES conductivity, particularly those including the effects of spatial dispersion and viscosity. These effects in 2DES are important when the frequency of electron-electron collisions is comparable to the radiation frequency~\cite{Conti_elasticity,Crossover}. These two effects are expected to suppress the excitation of high-oder modes.

This work was supported by Grants No. 16-29-03402 and 16-37-60110/16 of the Russian Foundation for Basic research.

\bibliography{Bibliography}

\end{document}